\documentclass[sigconf]{acmart}

\usepackage{bm}
\usepackage{booktabs}
\usepackage{dirtytalk}
\usepackage{makecell}
\usepackage{multirow}

\AtBeginDocument{%
  \providecommand\BibTeX{{%
    \normalfont B\kern-0.5em{\scshape i\kern-0.25em b}\kern-0.8em\TeX}}}

\copyrightyear{2020} 
\acmYear{2020} 
\setcopyright{rightsretained} 
\acmConference[KDD '20]{Proceedings of the 26th ACM SIGKDD Conference on Knowledge Discovery and Data Mining}{August 23--27, 2020}{Virtual Event, CA, USA}
\acmBooktitle{Proceedings of the 26th ACM SIGKDD Conference on Knowledge Discovery and Data Mining (KDD '20), August 23--27, 2020, Virtual Event, CA, USA}
\acmDOI{10.1145/3394486.3403046}
\acmISBN{978-1-4503-7998-4/20/08}

\begin{document}
% \fancyhead{}

\title{Preserving Dynamic Attention for Long-Term \\ Spatial-Temporal Prediction}

\settopmatter{authorsperrow=4}

\author{Haoxing Lin}
\authornote{Both authors contributed collectively to this research with a division of work. Haoxing Lin wrote this paper and designed the proposed algorithm and the experiments. Rufan Bai investigated the current methods and evaluated the baselines and variants. They worked together to run the experiments and collect the results.}
\orcid{0000-0001-9594-1871}
\author{Rufan Bai}
\authornotemark[1]
\affiliation{%
  \institution{State Key Lab of IoTSC\\
  FST, University of Macau}
}
\email{starklin96@gmail.com}
\email{yb97439@um.edu.mo}

\author{Weijia Jia}
\authornote{Contact Author}
\affiliation{%
  \institution{Joint AI and Future Network Research Institute, BNU (Zhuhai) \& UIC\\
  IoTSC, University of Macau}
}
\email{jiawj@sjtu.edu.cn}

\author{Xinyu Yang}
\affiliation{%
  \institution{State Key Lab of IoTSC\\
  FST, University of Macau}
}
\email{mb95466@um.edu.mo}

\author{Yongjian You}
\affiliation{%
  \institution{Shanghai Jiaotong University}
}
\email{youyongjian@sjtu.edu.cn}

\renewcommand{\shortauthors}{Lin and Bai, et al.}

\begin{abstract}
  Effective long-term predictions have been increasingly demanded in urban-wise data mining systems. Many practical applications, such as accident prevention and resource pre-allocation, require an extended period for preparation. However, challenges come as long-term prediction is highly error-sensitive, which becomes more critical when predicting urban-wise phenomena with complicated and dynamic spatial-temporal correlation. Specifically, since the amount of valuable correlation is limited, enormous irrelevant features introduce noises that trigger increased prediction errors. Besides, after each time step, the errors can traverse through the correlations and reach the spatial-temporal positions in every future prediction, leading to significant error propagation. To address these issues, we propose a Dynamic Switch-Attention Network (DSAN) with a novel Multi-Space Attention (MSA) mechanism that measures the correlations between inputs and outputs explicitly. To filter out irrelevant noises and alleviate the error propagation, DSAN dynamically extracts valuable information by applying self-attention over the noisy input and bridges each output directly to the purified inputs via implementing a switch-attention mechanism. Through extensive experiments on two spatial-temporal prediction tasks, we demonstrate the superior advantage of DSAN in both short-term and long-term predictions. The source code can be obtained from https://github.com/hxstarklin/DSAN.
\end{abstract}

  


\maketitle

\section{Introduction}
Predicting long-term futures has become one of the most urgent demands for urban-computing systems. More and more city-wise operations require several hours for preparation before the final execution, e.g., dynamic traffic management and intelligent service allocation \cite{WuT16}. Compared to traditional time-series problems, predicting spatial-temporal phenomena is more challenging because it has to handle not only the non-linear temporal correlation but also the dynamic and complicated spatial correlation. The challenge becomes more formidable in the long-term prediction as a small error can traverse through the complicated correlations, leading to the butterfly effect of error propagation that undermines the upcoming prediction of every spatial-temporal position. Currently, how to achieve effective long-term prediction of spatial-temporal phenomena is still a great challenge in data mining and machine learning communities.
 
The formal definition of the long-term prediction problem can be specified as, given the historical observations of a spatial-temporal phenomenon, learning a function that maps the inputs to the corresponding outputs of multiple future time steps. Recently, deep neural networks have been increasingly investigated for spatial-temporal prediction and outperform traditional approaches by employing sophisticated architectures. For example, deep residual Convolutional Neural Network (CNN) \cite{DeepResidual} has demonstrated outstanding performance in measuring the dynamic and complicated spatial correlations \cite{stresnet,ZHANG_TKDE,deepstd}. At the same time, Recurrent Neural Networks (RNN), especially the Long Short-Term Memory (LSTM) network \cite{lstm}, are frequently investigated for modeling the erratic temporal correlations \cite{lstmnn,DBLP:cui_ke_wang,deeptransport}. In the latest studies, convolutional recurrent structures -- the hybrid architectures that combine CNN and RNN, are proposed to jointly model the complicated and dynamic spatial-temporal correlations \cite{ConvLSTM,dmvstnet,stdn}. However, we observe that most of the current methods only focus on short-term prediction and become less effective in long-term prediction. One reason is that, in the historical observation, a considerable amount of irrelevant spatial information introduces noises into the prediction, which is mostly overlooked and leads to soaring errors. Moreover, by propagating through the complicated correlations, the errors from the previous outputs are interfering with the upcoming predictions when generating long-term results.

\begin{figure}[ht]
  \centering
  \includegraphics[width=0.85\linewidth]{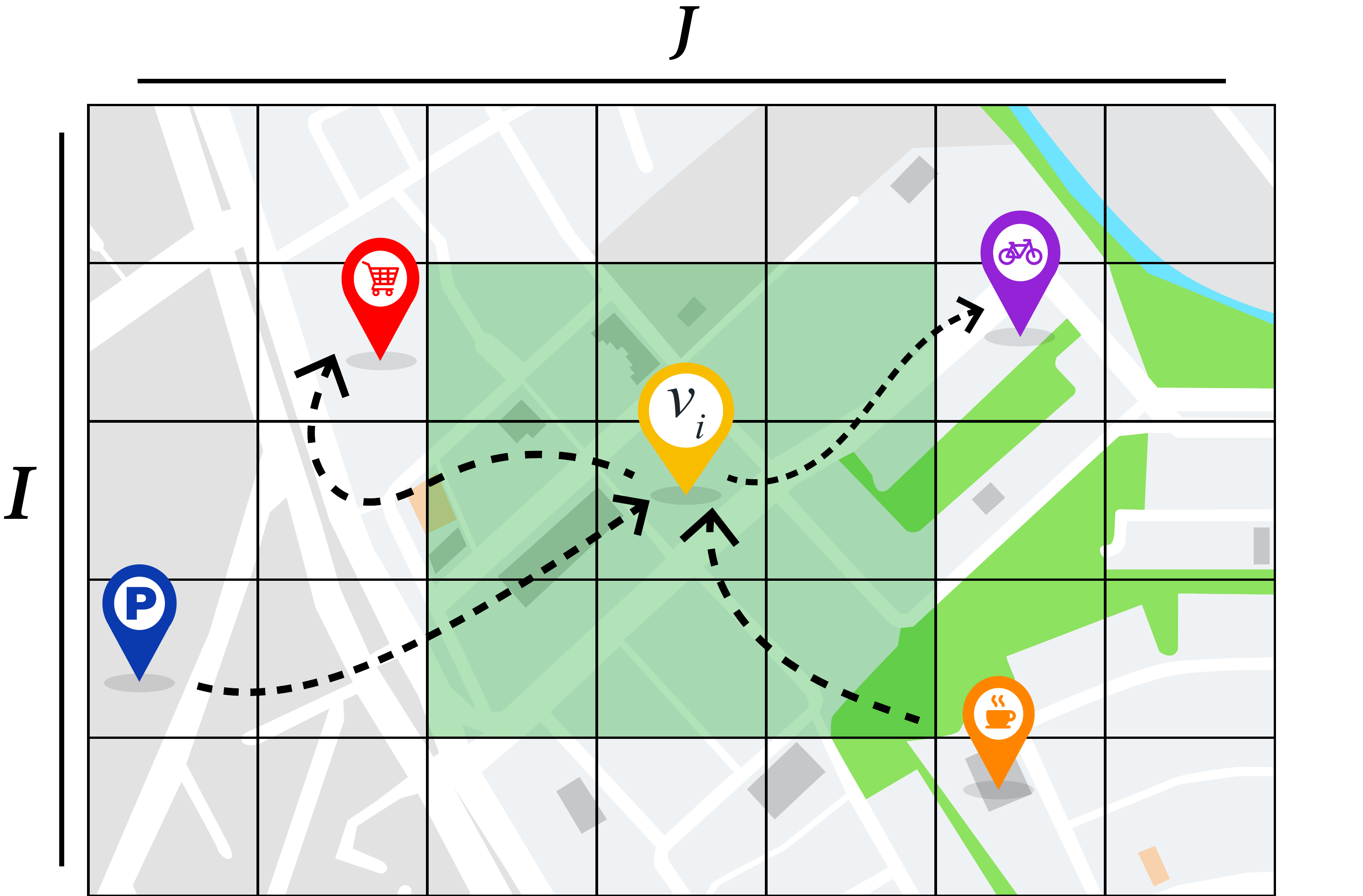}
  \caption{Not all spatial information is necessary for the prediction of $v_i$. Usually, only the closest neighbors (green) and those distant but unique locations (pined) are correlated with $v_i$. (For clarity, the figure is zoomed into a small area, which is different from the actual sampling resolution.)}
  \Description{fig_probdef}
  \label{fig_probdef}
\end{figure}

While evaluating the current methods, we observe the negative effect of irrelevant information as the error rates increase considerably when the spatial sizes of inputs become larger. In other words, not all spatial positions contain information that is helpful for the prediction (Figure~\ref{fig_probdef}). Instead, a larger spatial area is considered, the higher proportion of the input is trivial, introducing noises that further undermine the prediction. A common strategy to alleviate this effect is to limit the sampling area inside a local block, such that all inputs share strong correlations with the target \cite{stdn}. However, since the correlations are dynamic and distributed irregularly, considering only the closest neighbors also sacrifices the inputs that are distant but nontrivial for the prediction \cite{st_mgcn}.
 
We also observe that, following the traditional time-series prediction fashion, long-term predictions can easily absorb the errors from the previous results. When every spatial-temporal output passes its error to the next prediction, the error propagation is more significant than in typical time-series problems. Therefore, as the current methods focus on predicting the next time step only, they overlook this issue and no longer maintain their effectiveness in long-term predictions.
 
To address these issues, we propose a novel neural network named \textbf{D}ynamic \textbf{S}witch-\textbf{A}ttention \textbf{N}etwork (DSAN) with a Multi-Space Attention (MSA) mechanism to measure the complicated and dynamic correlation explicitly. Based on the MSA, we design two frameworks in the DSAN, including a Dynamic Attention Encoder (DAE) and a Switch-Attention Decoder (SAD), to filter out irrelevant noises and alleviate the error propagation. Employing a dual-encoder structure, the DAE performs self-attention over a global input and a strong-correlated input, then extracts the nontrivial features into a condensed representation through a regular attention mechanism. In the SAD, we develop a switch-attention structure with two decoders, which directly connects the output of each future time step to the inputs in different attention subspaces.
 
In summary, we make the following contributions in this paper:
\begin{itemize}
  \item We study the long-term spatial-temporal prediction problem and discover that filtering out irrelevant information with a dynamic distribution and alleviating error propagation are critical for achieving reliable performance.
  \item We propose the MSA that relies entirely on attention mechanisms to measure the spatial-temporal correlations by directly relating each position of the input and output.
  \item We propose the DSAN that consists of the DAE and SAD to filter out irrelevant noises in a global input and prevent significant error propagation.
  \item We extensively evaluate DSAN on two different tasks using three real-world data sets and demonstrate that the DSAN outperforms seven competing baselines by reducing at least 5\% of the RMSEs at each of the next 12 time steps (6 hours) on the Taxi-NYC data set.
\end{itemize}

\begin{figure*}[t]
  \centering
  \includegraphics[width=0.74\textwidth]{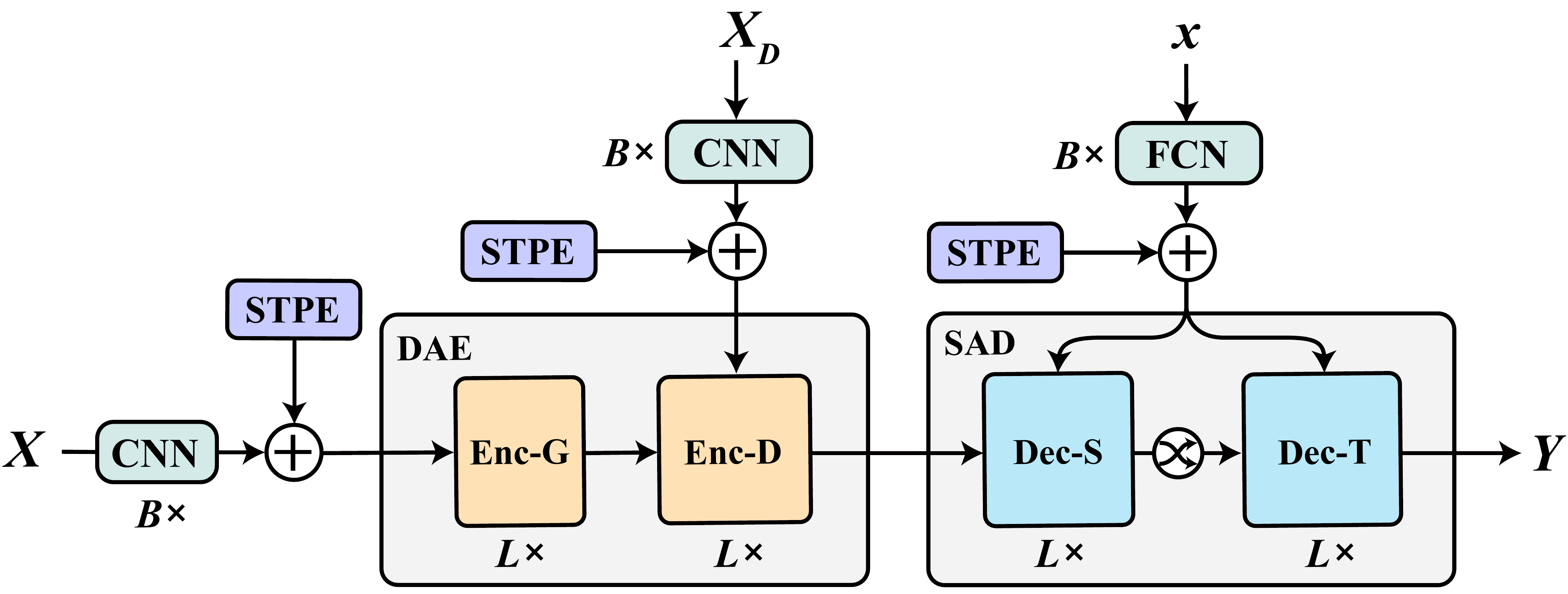}
  \caption{DSAN model architecture. $\bm{X}_D$ is a subset of $\bm{X}$, of which the grids are located in a local block surrounding $v_i$. $\bm{x}$ is the current features of $v_i$ extracted from $\bm{X}$. For the STPEs, their calculations of spatial positional encodings are different according to $\bm{X}$, $\bm{X}_D$, and $\bm{x}$. For clarity, we omit the details here.}
  \Description{fig_arch}
  \label{fig_arch}
\end{figure*}

\section{Related Work}
\subsection{Deep learning for spatial-temporal prediction}
Recently, employing deep neural networks has achieved significant improvement in spatial-temporal prediction. Since LSTM has been firmly established as a state of the art method for time-series prediction, it is widely employed to improve the performance of traffic state prediction \cite{DBLP:cui_ke_wang}. Later, noticing that not only temporal information but also the spatial correlations are critical, the surrounding areas' features are also considered in traffic flow prediction \cite{Zhang:2016:DPM:2996913.2997016}. From then on, a lot of spatial-temporal prediction researches, including predicting crowd flow \cite{stresnet} and ride-hailing demand \cite{ke_zheng_yang}, started to use CNN as the major tool for measuring spatial correlations. Demonstrating impressive effectiveness, deep residual CNNs \cite{stresnet,ZHANG_TKDE} and convolutional recurrent networks \cite{ConvLSTM,shi_2017,dmvstnet,stdn} are extensively investigated for jointly capturing the spatial-temporal correlations. In the latest works, graph convolutional network \cite{st_mgcn} is combined with LSTM to enhance the spatial-temporal measurements by leveraging graph representation. However, they overlook the negative effect introduced by irrelevant spatial information, which leads to increased errors when the considered area grows larger. Besides, they mainly focus on short-term prediction and pay little attention to the error propagation, which makes them less effective while predicting further results in the future.
  
\subsection{Graph-structured spatial-temporal prediction}
Some other spatial-temporal prediction researches are based on graph-structured data (e.g., highway sensor network data), which has been increasingly investigated as well. By leveraging graph convolution \cite{gcn,deep_gcn}, several works achieve effective measurement of the spatial features in graphs. For instance, DCGRU \cite{dcgru} and LC-RNN \cite{lcrnn} achieved remarkable improvements in capturing the local spatial correlations on traffic networks. ST-GCN utilizes multiple graph convolutional structures in traffic forecasting to simultaneously measure the spatial and temporal correlations \cite{stgcn}. GSTNet continues to improve traffic state prediction by employing sophisticated measurements of global dynamic correlations \cite{GSTNet}. LRGCN introduces R-GCN with Long Short-Term Memory and a novel path embedding method to obtain the state of the art performance in path failure prediction \cite{lrgcn}. GMAN implements a graph multi-attention structure to explicitly extract the relationships inside spatial and temporal dimensions and achieve consecutive prediction \cite{gman}. As these methods take as input graph-structured information, their effectiveness is limited by the graph neural networks, which has difficulty increasing the number of layers due to rank vanishing \cite{msdgcn}. Besides, the range of long-term prediction for graph-based methods is usually within an hour \cite{wavenet}.

\section{Notation and Problem Formulation}
As shown in Figure~\ref{fig_probdef}, we partition the spatial area into $N = I \times J$ grids and denote them as $\{v_1, v_2, ..., v_N\}$. The actual grid size empirically varies from $0.1km \times 0.1km$ to $1km \times 1km$ according to the quality and quantity of data. Each grid includes $b$ types of features, e.g., inflow and outflow in crowd flow prediction; service duration and request number in service utilization prediction. We use $X^t = \{x_1^t, x_2^t, ..., x_N^t\}$ to denote the historical observations of all grids at time $t$, where $x_n^t \in \mathbb{R}^b$ holds the observed features of $v_n$. When predicting the futures of a grid $v_i$, three inputs -- $\bm{X}_i$, $[x,y]_i$, and $\bm{r}_i$, are used to generate an output $\bm{Y}_i$. Here, $\bm{X}_i = \{X^{t_1}, X^{t_2}, ..., X^{t_h}\}$ is the historical spatial-temporal input, $[x,y]_i$ is the coordinate of $v_i$, and $\bm{r}_i = \{r_i^{t_1}, r_i^{t_2}, ..., r_i^{t_h}\}$ is the external input, where $r_i^t$ is a vector that consists of external features at time $t$. The details of the coordinate and the external information will be illustrated in Section~\ref{STPE}. The output $\bm{Y}_i = \{Y_i^1, Y_i^2, ..., Y_i^F\}$ is a sequence of results in the next $F$ time steps, where $Y^f \in \mathbb{R}^b$ contains $b$ predicted features of $v_i$ at the $f$-th upcoming time step. When predicting the results of $v_i$ and $v_j$ ($i \neq j$) at the same future time step, we have $\bm{X}_i = \bm{X}_j$ and $\bm{r}_i = \bm{r}_j$, but $[x,y]_i \neq [x,y]_j$. Notice that, in $\bm{X}_i$ and $\bm{r}_i$, the $\{t_1, t_2, ..., t_h\}$ does not have to be a sequence of consecutive time steps in the past, observations in one week ago and several days ago can also be included. In summary, the long-term prediction task can be formulated as learning a function $f_\theta(\cdot)$ that generate $\bm{Y}_k$ based on the inputs $\bm{X}_k$, $[x,y]_k$ and $\bm{r}_k$:
  
\begin{equation}
  \bm{Y}_k = f_\theta(\bm{X}_k, [x,y]_k, \bm{r}_k)
\end{equation}
 
\noindent where $k$ is the index of a training instance, $\bm{Y}_k = \{Y_k^1, Y_k^2, ..., Y_k^F\}$ are the predicted outputs and $\theta$ denotes the learnable parameters. Our goal is to minimize $\mathcal{L}$, the weighted mean square error, over the training set $D$:
 
\begin{equation}
  \mathop{{\rm \arg\min}}_{\theta} \mathcal{L} = \frac{\sum_{\{\bm{X}_k, [x,y]_k, \bm{r}_k\} \in D} \sum_{t=1}^{F} w_t (\hat{Y}_k^t - Y_k^t)^2}{|D|}
\end{equation}
 
\noindent where $\bm{X}_k$, $[x,y]_k$, and $\bm{r}_k$ are the inputs of a training instance, $\hat{\bm{Y}}_k =$\linebreak $\{\hat{Y}_k^1, \hat{Y}_k^2, ..., \hat{Y}_k^F\}$ are the corresponding ground truths of the next $F$ time steps, and $w_t$ denotes the weight of the $t$-th upcoming time steps in the loss function for joint training. The notations are listed in Appendix~\ref{apx_nota}.

\section{Methodology}
The model architecture of DSAN is demonstrated in Figure~\ref{fig_arch}. The inputs of DSAN are projected into $d$-dimension representations using $B$-layer fully-connected networks (FCN) or CNNs. They are summed with their corresponding Spatial-Temporal Positional Encodings (STPE) before entering DAE or SAD. The DAE is designed as a dual-encoder framework that contains two $L$-layer encoders -- Enc-G and Enc-D. Similarly, the SAD implements two $L$-layer decoders -- Dec-S and Dec-T, between which a switching operation is performed. Both DAE and SAD constructs each of their layers using the MSA, followed by a point-wise feed-forward network (FFN). We first detail MSA and STPE in Section~\ref{MSTA}, then elaborate DAE and SAD in Section~\ref{DAE} and~\ref{SAD}.
  
\subsection{Modeling Spatial-Temporal Attention} \label{MSTA}
Attention mechanism is widely used in natural language processing and computer vision to capture the relationships among words and pixels \cite{struct_att,struct_self_att}. Relying entirely on attention mechanism has achieved significant improvement in measuring sophisticated correlations by directly relating each position of the inputs and the outputs \cite{transformer,bert,gpt2,xlnet}, which could be beneficial for measuring spatial-temporal correlation. However, we observe that the amount of spatial-temporal input usually exceeds the effective resolution of attention computation \cite{transformer_xl}. Furthermore, the attention mechanism is less effective if no indication of the spatial and temporal difference between inputs is given. To solve these problems, we propose the MSA that separates the measurement of spatial-temporal attention into different subspaces and the STPE to indicate each input's unique identity.
  
\subsubsection{Multi-space attention.} \label{MSA}
Inspired by multi-head attention \cite{transformer}, MSA truncates the spatial-temporal information into different subspaces and separates the attention computation. It prevents over-averaging of attention weight after the $softmax$ function, albeit at the cost of leaving the results in different subspaces, which we counteract with the switch-attention mechanism introduced in Section~\ref{SAD}.
  
Taking $Q \in \mathbb{R}^{h \times L_q \times d}$, $K \in \mathbb{R}^{h \times L_k \times d}$, and $V \in \mathbb{R}^{h \times L_k \times d}$ as inputs, MSA calculates the attention-weighted outputs via scaled dot-product function. Here, $Q$, $K$, and $V$ refer to query, key, and value as in the typical attention mechanism \cite{NMTJL} and are acquired from different sources regarding the position of MSA. The number and size of the subspace are denoted as $h$ and $L_q/L_k$, e.g., the numbers of historical time steps and their spatial grids. As we also implement the multi-head attention in MSA, the attention mechanism performed by the $i$-th attention head can be formulated as:
  
\begin{equation}
  Att(Q_i, K_i, V_i, M) = softmax(\frac{Q_iK_i'}{\sqrt{d_h}} + M)V_i
\end{equation}
  
\noindent where the transpose of $K_i$ is performed over the last two axes, and $d_h = d/n_h$ denotes the split dimensionality in $n_h$ attention heads. $M \in \mathbb{R}^{h \times L_q \times L_k}$ is an artificial mask that adjusts the inputs of the $softmax$ function to manipulate the attention scores, helping to filter out certainly trivial grids. The calculation of $M$ is different according to the inputs and will be specified in the following subsections. The formula of MSA can be summarized as:
  
\begin{equation}
  MSA(Q, K, V, M) = \big[||_{i = 1}^{n_h} Att(QW_i^q, KW_i^k, VW_i^v, M)\big]W^o
  \label{eq_msa}
\end{equation}
  
\noindent where $||$ indicates the concatenation along the last axis and $W_i^q \in \mathbb{R}^{d \times d_h},\ W_i^k \in \mathbb{R}^{d \times d_h},\ W_i^v \in \mathbb{R}^{d \times d_h}$, and $W^o \in \mathbb{R}^{d \times d}$ are learnable projection matrices.

Notice that the axes of the inputs $Q,\ K$, and $V$ can be switched from $\mathbb{R}^{h \times L^\star \times d}$ to $\mathbb{R}^{L^\star \times h \times d}$ to further performed the attention calculation across different subspaces.
  
\subsubsection{Spatial-temporal positional encoding.} \label{STPE}
Since the MSA attends to the spatial-temporal information concurrently and equivalently, it does not know their relative positions and time differences. Like a person living in a world without color, it becomes difficult to focus on valuable objects. To solve this problem, the STPE is proposed as a particular bias to indicate the locations and time information.

\begin{figure}[t]
  \centering
  \includegraphics[width=0.88\linewidth]{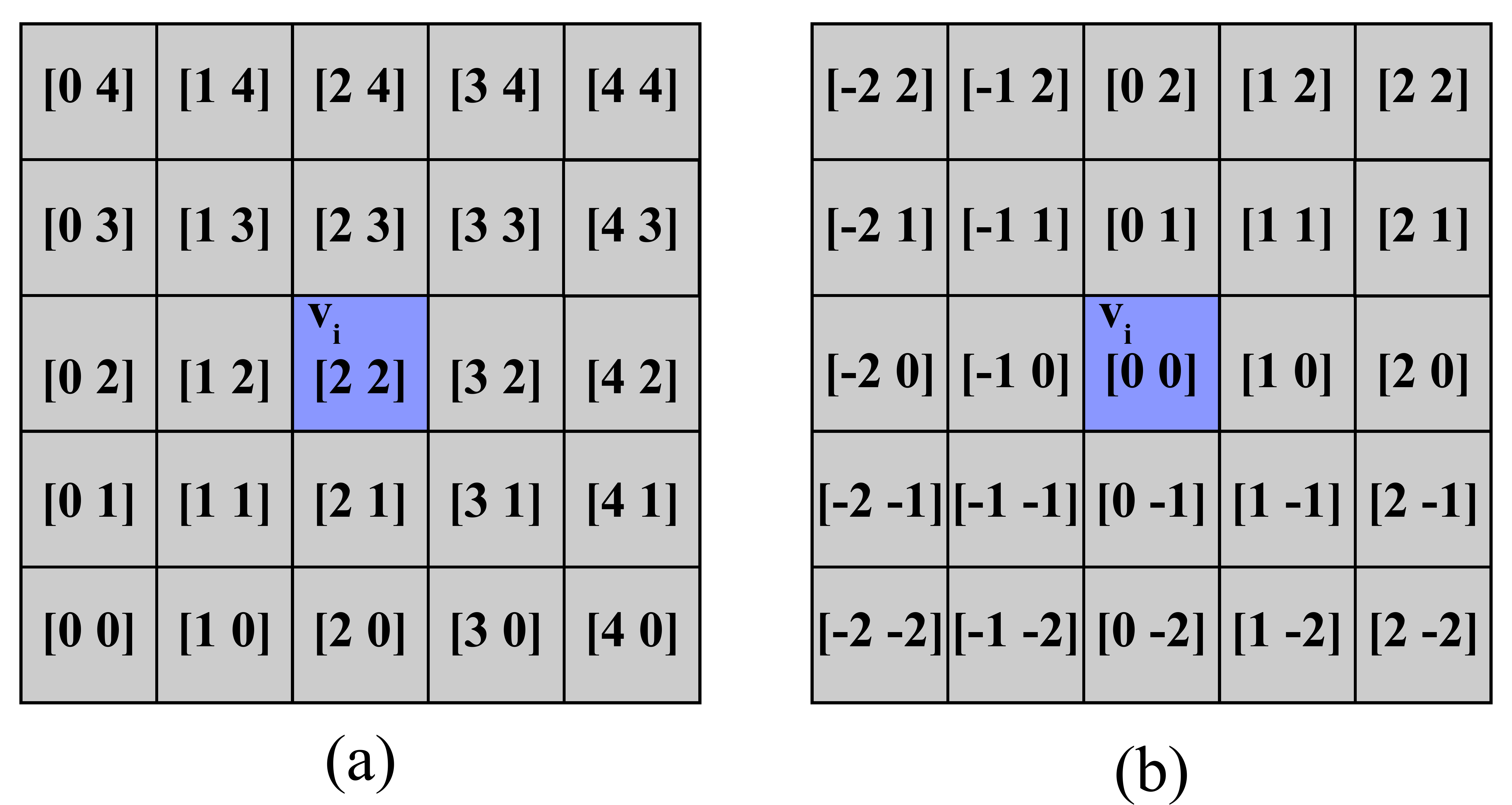}
  \caption{coordinate matrix: (a) absolute; (b) relative.}
  \Description{fig_spe}
  \label{fig_spe}
\end{figure}

In the STPE, we first calculate the Spatial Positional Encoding (SPE) for each grid. As shown in Figure~\ref{fig_spe}, a coordinate matrix is calculated according to $[x,y]_i$. We adopt two schemes for the calculation, including an absolute scheme that uses the original coordinates (Figure~\ref{fig_spe} (a)) and a relative scheme that recalculates all coordinates based on their relative positions to $[x,y]_i$ (Figure~\ref{fig_spe} (b)). Empirically, we observe that the relative scheme achieves better performance (Section~\ref{eval_variants}). One reason is that the coordinates around different $v_i$ can remain static and make the training more stable.
  
We calculate each dimension of the SPE given the coordinate matrix using sine and cosine functions of different frequency:
  
\begin{equation}
  SPE_{r,c}^l = \begin{cases}
  &sin(r/10000^{2l/d})\ \ \ \ \mathrm{if}\ l = 2n \\
  &cos(c/10000^{2l/d})\ \ \ \ \mathrm{if}\ l = 2n + 1
  \end{cases}
\end{equation}
  
\noindent where $SPE_{r,c} \in \mathbb{R}^{d}$ is the encoding vector of position $[r,c]$ in the matrix and $l$ indicates the $l$-th dimension. This mimics the positional encoding for language sequence \cite{transformer} and extends the representation to cover spatial-temporal information. After the calculations of all dimensions, the SPE matrix $SPE \in \mathbb{R}^{1 \times N \times d}$ is formed. Notice that, given a specific $v_i$, the $SPE$s for $\bm{X}$, $\bm{X}_D$, and $\bm{x}$ are static. They can be pre-calculated before the training.
  
Next, the Temporal Positional Encoding (TPE) is calculated to indicate the time information. First, from $\bm{r} \in \mathbb{R}^{h \times (7 + a + c)}$ we obtain the vector $r^t \in \mathbb{R}^{7 + a + c}$, whose first $7+a$ dimensions are the one-hot representation of the day-of-week and time-of-day information, and $a$ is the number of time step per day. The last $c$ dimensions are reserved for $c$ types of external information, including rainfall, snowfall, average temperature, and holiday. Then, we apply a two-layer fully connected network over $\bm{r} = \{r^1, r^2, ..., r^h\}$:
  
\begin{equation}
  TPE = \sigma(ReLU(\bm{r}W_1 + b_1)W_2 + b_2)
\end{equation}
  
\noindent where $W_1 \in \mathbb{R}^{(7 + a + c) \times d},\ W_2 \in \mathbb{R}^{d \times d},\ b_1,$ and $b_2$ are learnable parameters, $\sigma$ denotes the sigmoid activation, and $TPE \in \mathbb{R}^{h \times 1 \times d}$ is the TPE for $h$ historical time steps.
  
Finally, we sum the $SPE$ and the $TPE$ to generate $STPE \in \mathbb{R}^{h \times N \times d}$, where $SPE$ and $TPE$ are broadcasted to the same shape. The $STPE$s are summed with the projected $\bm{X}$, $\bm{X}_D$, and $\bm{x}$ to form the inputs of DAE and SAD.

\begin{figure}[t]
  \centering
  \includegraphics[width=0.72\linewidth]{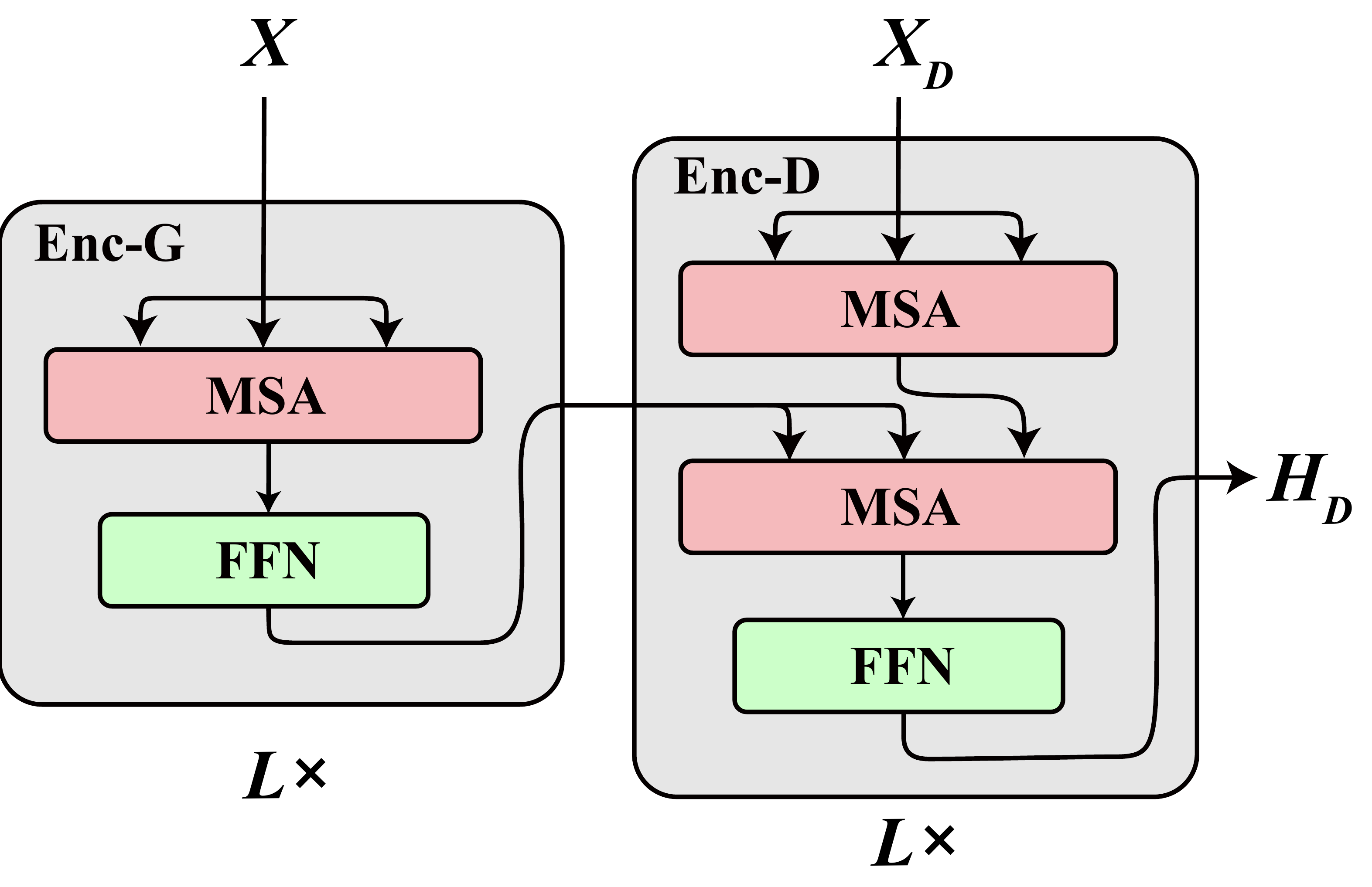}
  \caption{Dynamic attention encoder. Residual connections and layer-normalizations are omitted.}
  \Description{fig_dae}
  \label{fig_dae}
\end{figure}
  
\subsection{Dynamic Attention Encoder} \label{DAE}
During different predictions (e.g., different $v_i$'s, different time steps), the amount and distribution of nontrivial grids are dynamic. To minimize the impact of irrelevant noises, one can merely consider the closest neighbors within a local block. However, such a one-size-fits-all policy also sacrifices some distant but important grids. To address this problem, we propose the Dynamic Attention Encoder (DAE) to extract the valuable information that strongly correlates with $v_i$ from the global input. As shown in Figure~\ref{fig_dae}, the first encoder (Enc-G) calculates a representation of the global input using self-attention, and the second encoder (Enc-D) takes $\bm{X}_D$ as input to calculate a query $Q$ through self-attention as well. Then the $Q$ is used to select the nontrivial information from the global representation into a condensed representation through an additional MSA layer.
  
We use $\bm{H}_G^0 \in \mathbb{R}^{h \times N \times d}$ to denote the STPE-encoded input of historical observation $\bm{X}$. As shown in Figure~\ref{fig_dae}, each Enc-G layer performs self-attention MSA followed by a two-layer feed-forward network (FFN):
  
\begin{equation}
  \bm{P}_G^l = MSA(\bm{H}_G^{l-1}, \bm{H}_G^{l-1}, \bm{H}_G^{l-1}, M_G)
\end{equation}
  
\begin{equation}
  \bm{H}_G^l = FFN(\bm{P}_G^l) = ReLU(\bm{P}_G^lW_1^l + b_1^l)W_2^l + b_2^l
\end{equation}
  
In the self-attention layer, $Q,\ K$, and $V$ are obtained from the same source, which is the output of the previous layer. This allows positions in every subspace of $\bm{H}_G$ to correlate with each other. The FFN is used to perform linear transformation on each position separately and identically, where $W_1^l \in \mathbb{R}^{d \times d_f},\ W_2^l \in \mathbb{R}^{d_f \times d},\ b_1$ and $b_2$ are learnable parameters. We also use a threshold mask $M_G$ to filter out certainly useless (empty) grids, which is inspired by applying padding mask to erase invalid tokens \cite{bert}. We calculate the mask $M_G \in \mathbb{R}^{h \times N \times N}$ based on $\bm{X}$:
  
\begin{equation}
  M_G[t,:,i] = \begin{cases}
  &\ \ \ \ 0,\ \mathrm{if}\ \sum_{j=0}^{b-1} \bm{X}[t,i,j] > 0\\
  &-\infty,\ \mathrm{otherwise}
  \end{cases}
  \label{eq_mg}
\end{equation}
  
Consequently, the attention weights of trivial grids become zero in the outputs of the $softmax$ function (Equation~\ref{eq_msa}).
  
After Enc-G calculates the global representation $\bm{H}_G$, we implement the Enc-D to fulfill the dynamic extraction. As shown in Figure~\ref{fig_dae}, an Enc-D layer has an additional MSA in the middle, which calculates the weighted outputs from the $\bm{H}_G$ given a query $\bm{P}_D^l$ computed by the first MSA. We denote the initial input as $\bm{H}_D^0 \in \mathbb{R}^{h \times N_D \times d}$ ($N_D \le N$), which is similar to $\bm{H}_G^0$ but generated from $\bm{X}_D$. $\bm{X}_D$ is a subset of $\bm{X}$ that contains the closest neighbors that share strong correlations with $v_i$ within a local block. We use $L_d$ ($L_d = 2 l_d + 1$, where $l_d$ is a hyperparameter) to denote the length of the block and $N_D = L_d \times L_d$ is the number of the neighbors. If $v_i$ is a marginal grid, we use zero-padding to fill the vacant positions of the local block, which, compared to same-padding, empirically obtains a better performance. Given $\bm{H}_D^l$ and $\bm{H}_G$, the output of the Enc-D layer is calculated as:
  
\begin{equation}
  \bm{P}_D^l = MSA(\bm{H}_D^{l-1}, \bm{H}_D^{l-1}, \bm{H}_D^{l-1}, M_D)
\end{equation}
  
\begin{equation}
  \bm{H}_D^l = FFN\big(MSA(\bm{P}_D^l, \bm{H}_G, \bm{H}_G, M_G)\big)
\end{equation}
  
The calculation of $M_D \in \mathbb{R}^{h \times N_D \times N_D}$ is identical to $M_G$ as in Equation~\ref{eq_mg} but based on $\bm{X}_D$.
  
After passing through $L$ Enc-D layers, we finish the dynamic extraction of nontrivial information and generate the $\bm{H}_D \in \mathbb{R}^{h \times N_D \times d}$, which is later an input of the switch-attention decoder to generate the final output.

\begin{figure}[t]
  \centering
  \includegraphics[width=0.81\linewidth]{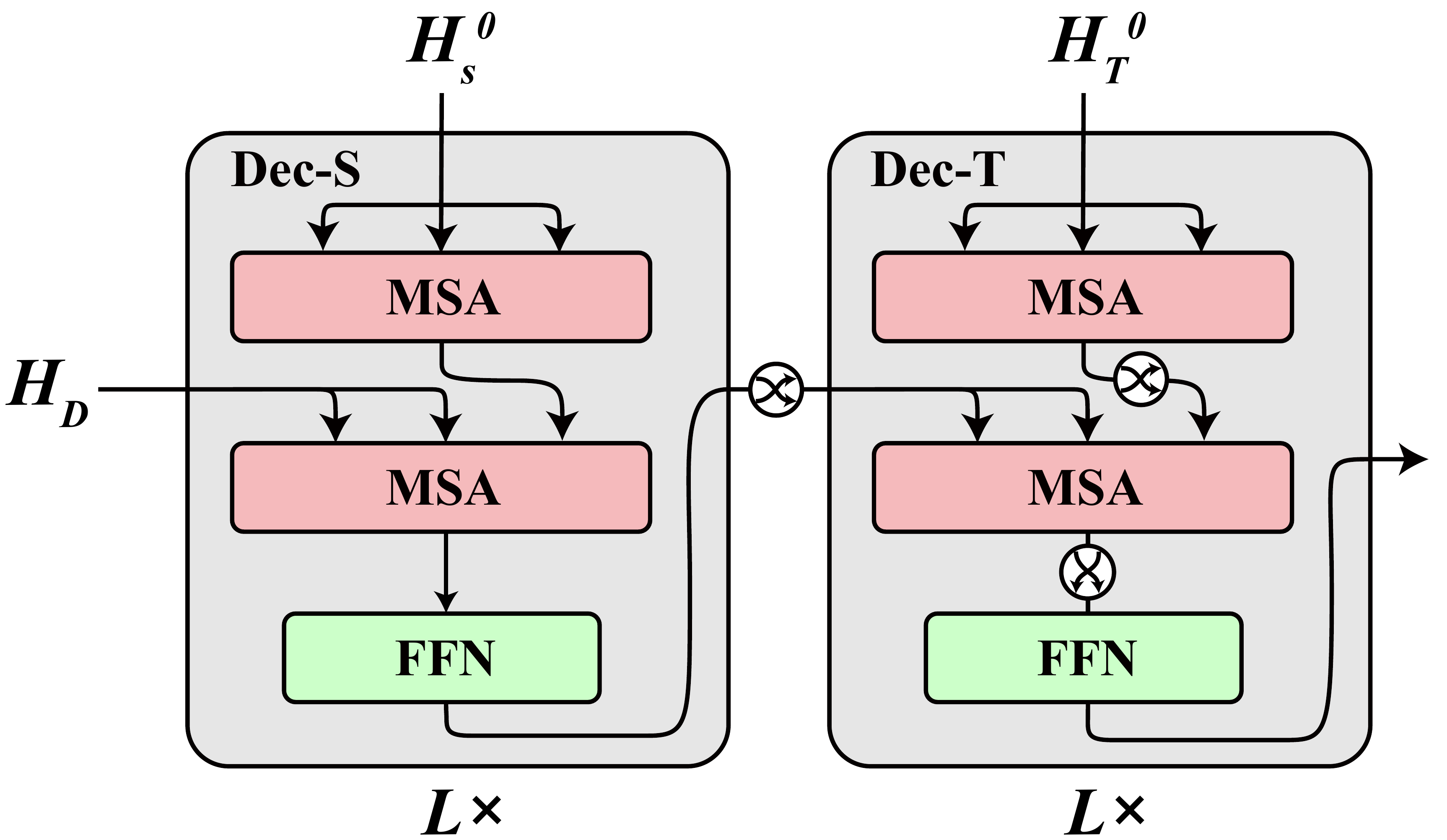}
  \caption{Switch-attention decoder. Residual connections and layer-normalizations are omitted.}
  \Description{fig_sad}
  \label{fig_sad}
\end{figure}

\begin{table*}[t]
  \centering
  \caption{Short-term prediction results (RMSE/MAPE[\%]). DSAN-ST only performs short-term prediction (1 future time step). DSAN-LT predicts 12 future time steps with equal joint training weights. DSAN-LTW predicts 12 time steps with 80\% joint training weight placed on the first time step. Other baselines perform short-term predictions only. Due to the space limitation, we only report the standard deviation of RMSE here.}
  \resizebox{\textwidth}{!}{
  \begin{tabular}{ l||c|c||c|c||c|c }
  \hline \hline
  \multirow{2}{*}{Model} & \multicolumn{2}{c||}{Taxi-NYC} & \multicolumn{2}{c||}{Bike-NYC} & \multicolumn{2}{c}{CTM} \\
  \cline{2-7}
  & inflow & outflow & inflow & outflow & duration (mins) & request \\
  \cline{2-7}
  \hline \hline
  MLP & 27.20$\pm$0.53/20.91 & 31.93$\pm$0.56/20.86 & 10.67$\pm$0.21/23.49 & 11.32$\pm$0.21/24.41 & 214.0$\pm$22/62.16 & 57.15$\pm$3.87/24.67 \\
  LSTM & 24.04$\pm$0.29/20.01 & 29.92$\pm$0.32/19.11 & 10.53$\pm$0.14/23.22 & 10.99$\pm$0.19/24.16 & 241.8$\pm$10/66.11 & 61.63$\pm$2.26/29.61 \\
  GRU & 23.36$\pm$0.30/18.51 & 29.77$\pm$0.25/19.31 & 10.66$\pm$0.15/23.12 & 11.35$\pm$0.16/24.25 & 219.5$\pm$16/53.44 & \textbf{53.88}$\pm$1.55/25.58 \\
  \hline
  ConvLSTM & 22.15$\pm$0.23/18.27 & 27.05$\pm$0.24/19.12 & 9.84$\pm$0.18/22.33 & 10.77$\pm$0.17/23.30 & 219.2$\pm$29/61.69 & 60.34$\pm$1.39/27.55 \\
  ST-ResNet & 20.65$\pm$0.25/18.02 & 25.52$\pm$0.27/18.93 & 9.28$\pm$0.14/21.72 & 10.39$\pm$0.13/22.79 & 216.1$\pm$19/58.38 & 58.25$\pm$1.26/25.31 \\
  DMVST-Net & 19.88$\pm$0.32/17.91 & 25.01$\pm$0.37/18.22 & 9.02$\pm$0.19/20.35 & 9.78$\pm$0.21/22.03 & 252.5$\pm$25/53.33 & 62.16$\pm$1.92/23.56 \\
  STDN & 19.46$\pm$0.21/17.57 & 24.49$\pm$0.33/17.28 & 8.67$\pm$0.10/19.46 & 9.43$\pm$0.11/20.31 & 294.3$\pm$28/96.75 & 66.38$\pm$1.53/22.30 \\
  \hline
  DSAN-LT & 18.01$\pm$0.55/16.83 & 23.75$\pm$0.61/17.38 & 9.07$\pm$0.23/19.58 & 9.36$\pm$0.22/20.21 & 188.7$\pm$25/52.28 & 64.42$\pm$2.03/13.45 \\
  DSAN-LTW & 17.95$\pm$0.31/16.21 & 23.38$\pm$0.32/16.20 & 7.37$\pm$0.18/18.57 & 8.47$\pm$0.18/19.45 & 178.5$\pm$28/\textbf{51.01} & 62.19$\pm$1.74/13.48 \\
  DSAN-ST & \textbf{17.65}$\pm$0.24/\textbf{16.05} & \textbf{22.86}$\pm$0.31/\textbf{16.10} & \textbf{7.23}$\pm$0.12/\textbf{18.39} & \textbf{8.46}$\pm$0.15/\textbf{19.23} & \textbf{177.1}$\pm$24/53.86 & 64.04$\pm$1.54/\textbf{13.38}\\
  \hline \hline
  \end{tabular}}
  \label{tbl_results_st}
\end{table*}
  
\subsection{Switch-Attention Decoder} \label{SAD}
As the MSA conducts attention calculation by truncating the spatial-temporal space, it solves the over-averaging problem but leaves separated results in different subspaces. Instead of performing linear transformation, which empirically results in poor performance (Section~\ref{eval_variants}), SAD switches its attention focus by transposing the outputs of DAE and further performs the MSA across subspaces to calculate the final output.
  
As shown in Figure~\ref{fig_sad}, the Dec-S and Dec-T layers both contain two MSAs and an FFN. The input $\bm{x} \in \mathbb{R}^{1 \times f \times b}$ is the feature sequence of $v_i$ where $f$ indicates the number of future time steps. During the evaluation and testing, $\bm{x}$ initially contains only the latest historical features of $v_i$ extracted from $\bm{X}$. It will be concatenated with new outputs after each time step to form the next input as in an autoregressive model \cite{NMTJL}. After the projection (FCN) and summation with STPE (Figure~\ref{fig_arch}), $\bm{x} \in \mathbb{R}^{1 \times f \times b}$ is broadcasted to $\bm{H}_S^0 \in \mathbb{R}^{h \times f \times d}$ as the input of Dec-S. We formulate the first MSA of the Dec-S layer as:
  
\begin{equation}
  \bm{P}_S^l = MSA(\bm{H}_S^{l-1}, \bm{H}_S^{l-1}, \bm{H}_S^{l-1}, M_z)
  \label{eq_decs_msa_1}
\end{equation}
  
\begin{equation}
  M_z[:,i,j] = \begin{cases}
  &\ \ \ \ 0,\ \mathrm{if}\ i \le j\ \mathrm{and}\ \sum_{k=0}^{b-1} \bm{x}[:,j,k] > 0\\
  &-\infty,\ \mathrm{otherwise}
  \end{cases}
\end{equation}
  
\noindent where $M_z \in \mathbb{R}^{h \times f \times f}$ is the combination of threshold mask and look ahead mask, which prevents each position of input to look into the future. In the rest part of Dec-S, given $\bm{P}_S^l$ as $Q$ and $\bm{H}_D$ as $K$ and $V$, we calculate the output as:
   
\begin{equation}
  \bm{H}_S^l = FFN\big(MSA(\bm{P}_S^l, \bm{H}_D, \bm{H}_D, 0)\big)
  \label{eq_decs_ffn}
\end{equation}
  
\noindent where $\bm{H}_S^l \in \mathbb{R}^{h \times f \times d}$ contains $f$ outputs calculated by the MSAs in $h$ subspaces. Notice that the mask is unnecessary here, as all irrelevant inputs are already filtered out in the previous computations.
  
Next, $\bm{H}_S$ (the output of Dec-S) and $\bm{H}_T^0 \in \mathbb{R}^{1 \times f \times d}$ (similar to $\bm{H}_S^0$ but without being broadcasted) are taken as inputs by Dec-T to generate the final output. Before entering Dec-T, the first two axes of $\bm{H}_S$ are switched to form $\bm{H}_S' \in \mathbb{R}^{f \times h \times d}$. After going through the first MSA, $\bm{P}_T^l$ is obtained in the same way as indicated in Equation~\ref{eq_decs_msa_1}. Then, inside a Dec-T layer, there are two switching gates applied on $\bm{P}_T^l$ and the second MSA's output over their first two axes (Figure~\ref{fig_sad}). Finally, $\bm{H}_T^l \in \mathbb{R}^{1 \times f \times d}$ is generated after the second MSA and the FFN as indicated in Equation~\ref{eq_decs_ffn}.
  
After Dec-T delivers its final output $\bm{H}_T$, we use a fully-connected network to generate the prediction $\bm{Y} \in \mathbb{R}^{f \times b}$:
  
\begin{equation}
  \bm{Y} = \sigma(\bm{H}_TW_y + b_y)
\end{equation}

\noindent where $W_y \in \mathbb{R}^{d \times b}$ and $b_y$ are the learnable parameters and $\sigma$ denotes the sigmoid activation.

\section{Experiments} \label{EXP}
We include two spatial-temporal prediction tasks in the Experiments: (1) crowd flow prediction; (2) service utilization prediction. Since the two tasks have different strengths of spatial-temporal correlation, they can verify the generalizations of DSAN and other baselines over different scenarios. Specifically, in crowd flow prediction, an outflow in a position corresponds to an inflow in another position, which indicates a strong correlation. However, the correlation is weaker and more difficult to measure in service utilization prediction, where the increase of service duration in one position may or may not indicates changes in other positions. The details of the data sets and the procedure of data preprocessing are shown in Appendix~\ref{apx_data}.
  
\subsection{Metrics \& Baselines} We compare DSAN with seven baselines based on two metrics: (1) Root Mean Square Error (RMSE) and (2) Mean Absolute Percentage Error (MAPE).
  
\subsubsection{Baselines.}
\begin{itemize}
  \item \textbf{(1) MLP}: Multi-Layer Perceptron, a three-layer fully-connected network.
  \item \textbf{(2) LSTM}: Long-short term memory.
  \item \textbf{(3) GRU}: Gated Recurrent Unit network \cite{DBLP:journals/corr/ChungGCB14}.
  \item \textbf{(4) ConvLSTM}: Convolutional LSTM \cite{ConvLSTM}, which extends the fully connected LSTM to have convolutional structures, is proposed for the precipitation nowcasting problem.
  \item \textbf{(5) ST-ResNet}: Spatial-Temporal Residual Convolutional Network \cite{stresnet}, which applies multiple deep residual convolutional networks to measure spatial correlations from different temporal periods to predict crowd flow.
  \item \textbf{(6) DMVST-Net}: Deep Multi-View Spatial-Temporal Network \cite{dmvstnet}, which considers three different views -- temporal view, spatial view, and semantic view in crowd/traffic flow prediction.
  \item \textbf{(7) STDN}: Spatial-Temporal Dynamic Network \cite{stdn}. A combination of CNN and LSTM that achieves remarkable improvement by considering the transition and temporal shifting in crowd flow prediction.
\end{itemize}

We obtained the means and standard deviations of RMSEs and MAPEs after running each method for ten times. The implementations of the baselines are detailed in Appendix~\ref{apx_baselines}. The hyperparameter setting of DSAN is illustrated in Appendix~\ref{apx_hyper}.

\begin{figure*}[t]
  \centering
  \includegraphics[width=0.9\textwidth]{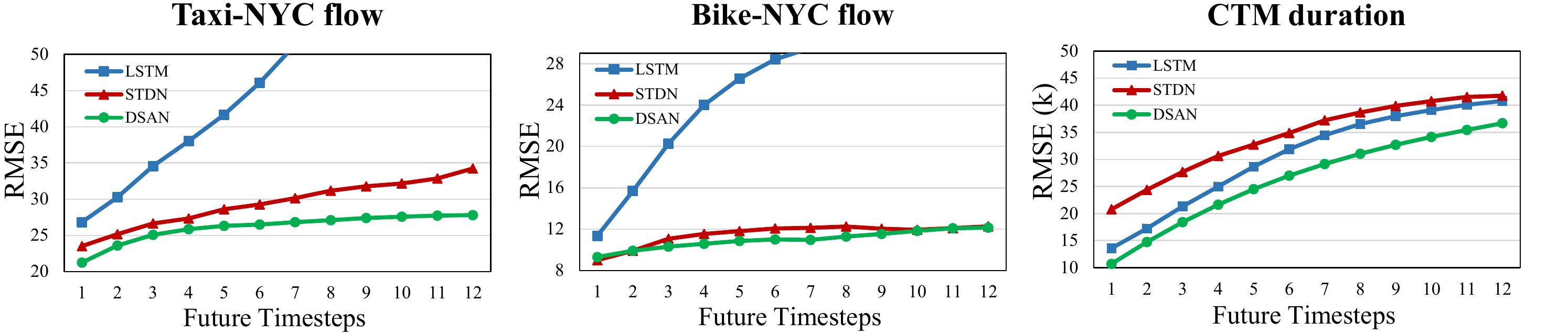}
  \caption{Long-term prediction results. The flows of Taxi-NYC and Bike-NYC are reported as the means of inflows and outflows.}
  \Description{fig_results_lt}
  \label{fig_results_lt}
\end{figure*}
  
\subsection{Results}
\subsubsection{Short-term prediction.}
In this subsection, we evaluate the effectiveness of DSAN and other baselines in modeling complicated and dynamic spatial-temporal correlations through inspecting the short-term prediction results. As shown in Table~\ref{tbl_results_st}, in the two crowd flow prediction tasks, the deep learning methods outperform the traditional neural networks given their sophisticated architectures dedicated to measuring spatial-temporal correlation simultaneously. However, as they rely on structures combining CNNs or RNNs, the problems facing these fundamental neural methods are also inherited. The most critical one is that the measurement of long-term correlation is far from effective due to the lengthy paths between outputs and distant inputs. For example, ConvLSTM focuses on short-term nowcasting by implementing LSTM in a convolutional style, which demonstrated impressive improvement in precipitation nowcasting but overlooked that the distant inputs have to traverse through the complicated structure, which makes it less effective when applied to measure long-term correlation. In comparison, ST-ResNet deploys multiple deep CNNs over spatial inputs from different periodic sequences and merges the results through linear transformation. Since it measures the spatial features in parallel, the long-term and short-term correlations share the same path length. Nevertheless, the non-linearity of the temporal correlations is ignored, which has unique impacts that should not be measured equivalently. The recent works addressed this issue by combining CNNs and LSTMs. For example, DMVST-Net captures the spatial features by CNNs and feeds the convolutional results to LSTMs. It also considers the semantic scenarios and achieved considerable improvement in crowd flow prediction. Similarly, STDN extends the advantage of DMVST-Net and achieves further improvement by including crowd flow transition through a gated mechanism and considering the periodic shifting of temporal patterns. However, they face a similar problem encountered by ConvLSTM. In comparison, through employing MSA, the output of DSAN can reach both long-term and short-term inputs explicitly through the unique attention links without passing through convolutional or recurrent structures. Besides, DSAN dynamically extracts the valuable features from the global inputs via DAE, which filters out irrelevant noises without sacrificing those distant but nontrivial grids. In comparison, other baselines reduce the amount of irrelevant information by limiting the spatial size of inputs to minimize the noise introduced, which ignores the nontrivial features that are distant from the target. Consequently, DSAN demonstrates reasonable improvement in the crowd flow prediction tasks.
  
In the service utilization prediction, the advantages of deep learning methods are not as distinguishable as in the crowd flow predictions. We observe that the fundamental neural network methods achieve decent performances compared to the deep learning methods. One reason is that the weak and erratic correlations among spatial-temporal positions in CTM data post greater challenges in achieving effective measurement. Besides, the deep learning baselines are majorly studied on measuring strong correlations, allowing traditional neural methods, especially the GRU discarding the memory unit, to obtain decent performance without considering the weak spatial correlations. Compared to other deep learning approaches, DSAN still outperforms most of the baselines. Nevertheless, it is profoundly affected by the erratic weak correlations and is not as competent as in the crowd flow prediction tasks.

\begin{table}[t]
  \centering
  \caption{Variants of DSAN on Taxi-NYC (short-term).}
  \begin{tabular}{ l||c|c }
  \hline \hline
  \multirow{2}{*}{Variants} & \multicolumn{2}{c}{RMSE/MAPE(\%)} \\
  \cline{2-3}
  & Inflow & Outflow \\
  \hline
  MHAN & 28.10/24.61 & 32.63/24.10 \\
  DSAN-NE & 23.96/20.43 & 29.80/20.02 \\
  DSAN-SD & 21.04/18.02 & 26.68/17.75 \\
  DSAN-SE & 18.83/16.77 & 24.17/16.22 \\
  DSAN-AE & 18.14/16.41 & 23.49/16.13 \\
  DSAN & \textbf{17.65}/\textbf{16.05} & \textbf{22.86}/\textbf{16.10} \\
  \hline \hline
  \end{tabular}
  \label{tbl_variants}
\end{table}
  
\subsubsection{Long-term prediction.} \label{Exp_LT}
In this subsection, we evaluate the performances of different methods, including LSTM, STDN, and DSAN, over predicting 12 future time steps. For STDN, as its original implementation performs the prediction of only one future time step, we extend the length of its last LSTM layer as well as the Bahadanau attention layer to allow a consecutive long-term prediction. When fulfilling the joint training, the weights of all time steps are equivalent for all evaluated methods.
 
As shown in Figure~\ref{fig_results_lt}, we observe the increased errors at the first time step for all evaluated approaches, compared to their results of merely performing short-term predictions. Specifically, on the Taxi-NYC data, DSAN has its RMSE increased by 3.5\%, while those of STDN and LSTM increase by 7\% and 18\%. At the further time steps, the RMSEs of STDN and LSTM increase by 5.3\% and 10.4\% averagely after each time step, while DSAN keeps it at a decent level with 3.4\%. It demonstrates that bridging every output to all inputs directly by the SAD is critical for alleviating the error propagation, compared to relying totally on the previously predicted outputs. Such an advantage is more evident as the long-term prediction goes further into the future, compared to the rising tails of the RMSE curves of LSTM and STDN (Figure~\ref{fig_results_lt}). Notice that DSAN maintains its advantage on the CTM data, where STDN is significantly less effective due to its dependency on the transition features for assisting the prediction of crowd flow. In comparison, DSAN can still perform a reliable spatial-temporal prediction without considering any assisting information.

\begin{table}[t]
  \centering
  \caption{RMSE (inflow/outflow) on Taxi-NYC given inputs with different spatial size. \textit{N/A} indicates that the DAE local block is larger than the input size. \textit{Global} indicates feeding the entire input without limiting the sampling area. The number after DSAN refers to the corresponding $l_d$. All variants perform short-term prediction only.}
  \resizebox{\linewidth}{!}{
  \begin{tabular}{ l||c|c|c|c }
  \hline \hline
  \multirow{2}{*}{Models} & \multicolumn{4}{c}{Spatial Size} \\
  \cline{2-5}
  & 5 $\times$ 5 & 9 $\times$ 9 & 13 $\times$ 13 & Global \\
  \hline
  STDN & 19.64/24.83 & 19.79/24.95 & 20.90/26.13 & 22.56/28.49 \\
  DSAN (2) & 19.25/24.30 & 18.74/23.91 & 18.80/24.09 & 18.74/23.85 \\
  DSAN (3) & N/A & 17.69/22.89 & 17.71/22.89 & \textbf{17.65}/\textbf{22.86} \\
  DSAN (4) & N/A & 18.08/23.27 & 18.02/23.31 & 17.93/23.13 \\
  DSAN (5) & N/A & N/A & 18.95/23.94 & 18.92/23.99 \\
  \hline \hline
  \end{tabular}}
  \label{tbl_eval_dae}
\end{table}
  
\subsubsection{Evaluation on model variants.} \label{eval_variants}
To demonstrate the effects of different components in DSAN, we evaluate the following variants:
 
\begin{itemize}
  \item MHAN: Multi-Head Attention Network. In this variant, we replace all MSAs with the Multi-Head Attention (MHA). When employing MHA to calculate the attention, the spatial-temporal inputs are flattened into a single space. Besides, only one decoder is used in MHAN since switch-attention is infeasible without the MSA. A fully-connected network is used to transform the decoder output to the final output.
  \item DSAN-NE: DSAN with No Encoding. The STPEs are not calculated in this variant, and the inputs directly go into DAE or SAD after the FCN/CNNs.
  \item DSAN-SE: DSAN with Single Encoder. Only one encoder is applied to the global input $\bm{X}$ while discarding $\bm{X}_D$. No dynamic extraction is performed.
  \item DSAN-SD: DSAN with Single Decoder. Only one decoder is employed without the switch-attention, and the output is flattened and transformed to the final result by a fully-connected network.
  \item DSAN-AE: DSAN with absolute spatial positional encoding.
\end{itemize}
  
As shown in Table~\ref{tbl_variants}, the variants are more or less not as competent as DSAN. Specifically, the RMSE increases by 60\% with MHAN, resulting in a result even worse than the one of MLP. The main reason is that spatial-temporal phenomena contain lots of inputs (more than one thousand in Taxi-NYC) that exceed the capability of attention calculation \cite{transformer_xl}. As a result, the attention weight of every position is over-averaged, and the generated attention scores are meaningless in MHAN. For other variants, without the STPE, DSAN-NE obtains a 35\% increase in RMSE as no indication about the spatial and temporal positions is given during the attention calculations. DSAN-SE, DSAN-SD, and DSAN-AE also produce higher RMSEs, which illustrates that the effectiveness of DSAN relies on the essential contributions delivered by DAE and SAD. 

\subsubsection{Evaluation on irrelevant noises.}
In this subsection, we evaluate feeding STDN and DSAN with different inputs to illustrate the impact of irrelevant noises and demonstrate the effectiveness of DAE. As shown in Table~\ref{tbl_eval_dae}, we obtain $\bm{X}$s with different $n \times n$ sampling blocks that surround $v_i$, where $Global$ indicates feeding the whole input as in the original setting. Notice that, the sampling block here is different from the local block mentioned in Section~\ref{DAE}, and it is only used to accomplish this evaluation. We also evaluate the effect of different $l_d$'s, which decide the size of DAE's local block for extracting a strongly-correlated query for Enc-D (Section~\ref{DAE}). As shown in the result, all DSAN variants maintain robust when the spatial size (and the volume of noise) grows. When $l_d = 2$, we observe that the performance keeps improving with a larger input as more valuable information is collected from distant positions. After setting $l_d$ to $3$, DSAN achieves the best performance when fed with the global input, demonstrating the effectiveness of selecting nontrivial spatial information dynamically. However, when $l_d$ goes up to $4$ or $5$, the local block itself contains a sufficient amount of noise, which deters further improvement even with larger inputs. In comparison, STDN achieves its best performance when the spatial size of the input is set between $5 \times 5$ and $9 \times 9$ ($7 \times 7$ is the best setting suggested by the authors) and has increasing errors when the input size grows. In summary, the irrelevant spatial noises play a key role in introducing higher errors in the predictions, and the dynamic extraction of nontrivial information fulfilled by the DAE is critical to secure a robust performance.

\begin{figure}[t]
  \centering
  \includegraphics[width=\linewidth]{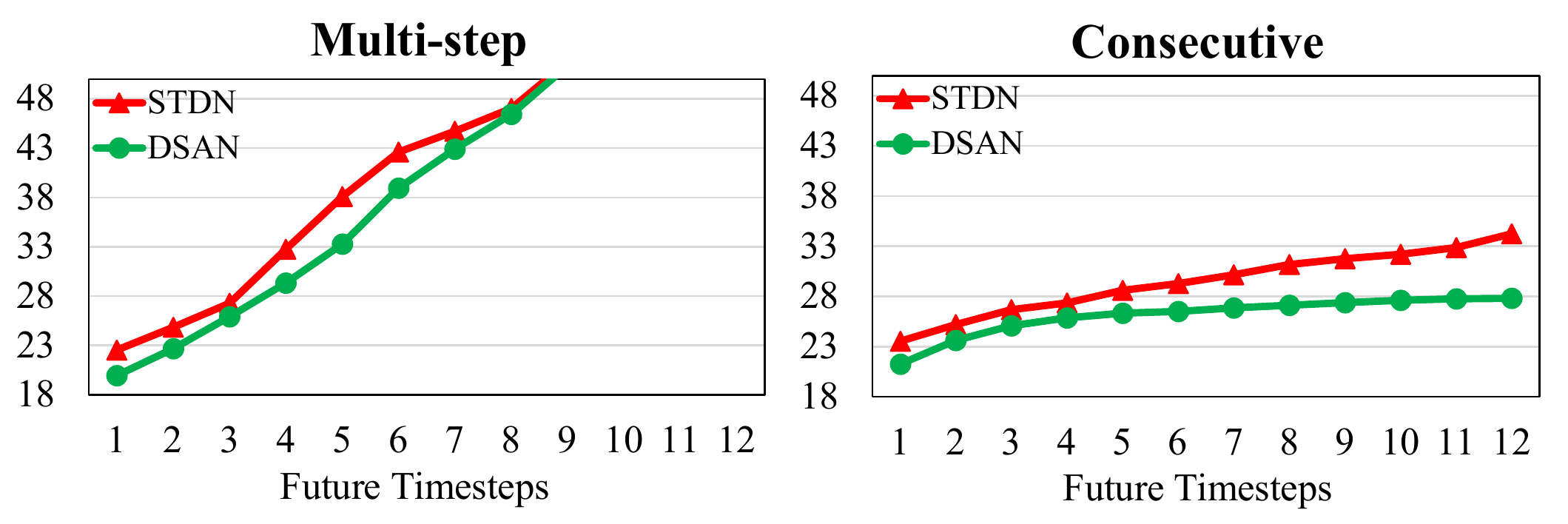}
  \caption{RMSE results on Taxi-NYC with different long-term strategies. Values are reported as the means of inflow and outflow.}
  \Description{fig_eval_ep}
  \label{fig_eval_ep}
\end{figure}

\subsubsection{Evaluation on long-term error propagation.}
To clearly illustrate how error propagation affects the long-term prediction, we further analyze the performances of STDN and DSAN based on two strategies: (1) multi-step prediction and (2) consecutive prediction. By adopting the multi-step strategy, the methods are trained on short-term prediction, which allows both STDN and DSAN to achieve their best predictions at the next time step. After finishing the predictions of all $v_i$'s, the whole spatial output is concatenated with the original input for the prediction at the next time step. In comparison, consecutive strategy trains both methods to predict all long-term results of $v_i$ consecutively in one prediction. It requires the methods to balance its power over different time steps, which usually makes the first prediction less accurate but alleviates the error propagation without absorbing the predicted errors from other $v_i$'s. Notice that, since STDN is implemented for short-term prediction, we modify its architecture to achieve the consecutive prediction (Section~\ref{Exp_LT}). As shown in Figure~\ref{fig_eval_ep}, despite both methods predict well at the first time step by adopting the multi-step strategy, the errors soar when predicting more future results. In comparison, the consecutive strategy leads to slightly higher errors in the first time step but significantly moderates the error increasing in the later predictions. Notice that, given both strategies, DSAN maintains its advantage over STDN, demonstrating the effectiveness of directly connecting the inputs and outputs by SAD.
  
\subsubsection{Evaluation on joint training.}
\begin{figure}[t]
  \centering
  \includegraphics[width=\linewidth]{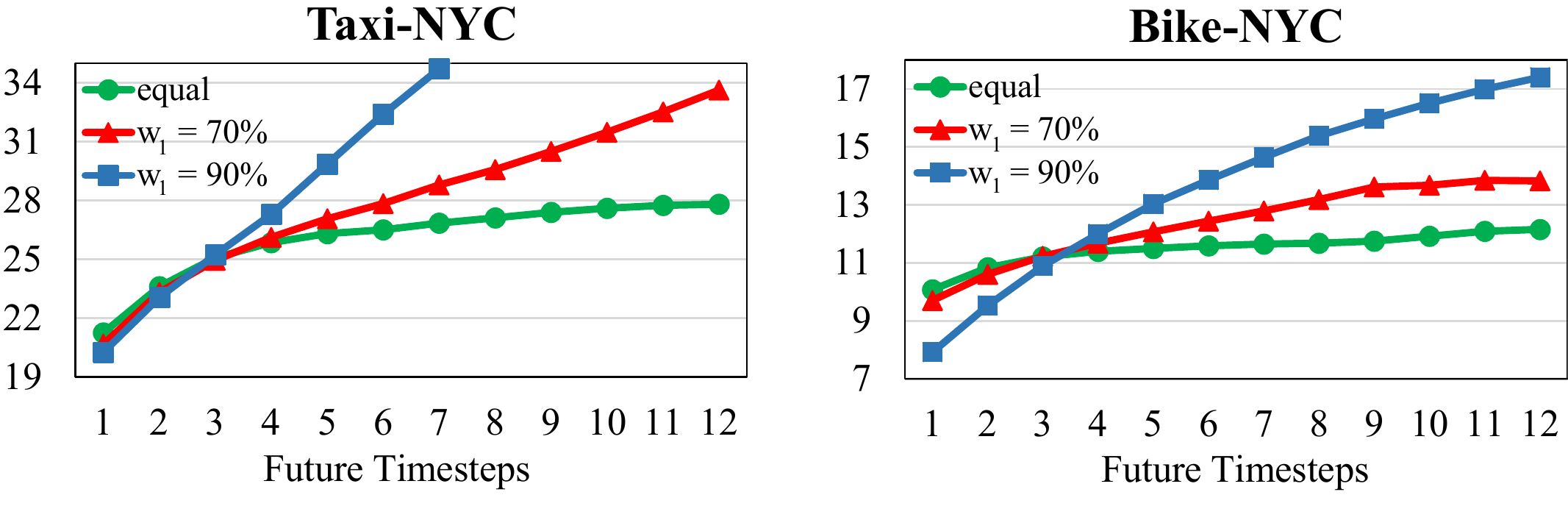}
  \caption{Joint training evaluation: RMSEs are the means of in/out flow.}
  \Description{fig_results_w}
  \label{fig_results_w}
\end{figure}

By adjusting the joint training weights of each time step, we can control the optimization during the training. As shown in Figure~\ref{fig_results_w}, when the weights are equally set, DSAN balances its power on every prediction. If the first time step is more important, giving it a larger weight can obtain considerable improvement, albeit at the cost of sacrificing the accuracies of other time steps. The two examples allocate 70\% and 90\% of training weight on the first time step, delivering more inclined RMSE curves with better predictions for the first time steps.

\section{Conclusion}
In this work, we study the impacts of irrelevant noises and error propagation in the long-term prediction of spatial-temporal phenomena and present a novel DSAN that achieves a new state of the art performance. With the MSA that utilizes truncation to measure the spatial-temporal attention, the DAE can dynamically extract nontrivial features while the SAD bridges every output directly to the inputs. However, the potential problem is that, as Smart City is developing rampantly, the increasing amount of spatial-temporal inputs may again exceed the capability of all existing methods. Trying to obtain further improvement in both short-term and long-term prediction is more computational unaffordable. Therefore, we believe that retrieving nontrivial information from a very large spatial-temporal space while maintaining a decent computational workload is the focus of our future works.

\begin{acks}
  This work is partially funded by The Science and Technology Development Fund, Macau SAR, FDCT.0007/2018/A1, FDCT.0060/2019/A1, University of Macau (File no. MYRG2018-00237-FST); Chinese National Research Fund (NSFC) Key Project No. 61532013 and No. 61872239. CTM Macau provided data for the concept of proof.
\end{acks}

\bibliographystyle{ACM-Reference-Format}
\bibliography{acmart}

\pagebreak
\appendix

\section{Notations} \label{apx_nota}
The notations in this paper are detailed in Table~\ref{tbl_nota}:

\begin{table}[ht]
  \centering
  \caption{Notation table.}
  \resizebox{\linewidth}{!}{
  \begingroup
  \renewcommand{\arraystretch}{1.618}
  \begin{tabular}{ l c }
  \toprule
  $v_1,\ ...,\ v_N$ & spatial grids\\
  $N$/$N_d$ & number of grids (Global/Enc-D)\\
  $b$ & number of features\\
  $t_1,\ ...,\ t_h$ & historical time steps\\
  $x_i^t$ & input of $v_i$ at time $t$\\
  $X^t = \{x_1^t,\ ...,\ x_N^t\}$ & all inputs at time $t$\\
  $\bm{X} = \{X^{t_1},\ ...,\ X^{t_h}\}$ & spatial-temporal inputs\\
  $\bm{X}, \bm{X}_D, \bm{x}$ & Enc-G, Enc-D, and SAD inputs\\
  $[x,y]_i$ & the coordinate of $v_i$\\
  $\bm{r} = \{r^{t_1},\ ...,\ r^{t_h}\}$ & external inputs\\
  $\bm{Y} = \{Y^1,\ ...,\ Y^F\}$ & predictions of next $F$ time steps\\
  $w_1,\ ...,\ w_F$ & joint training weights\\
  $d\ (d_h/d_f)$ & dimension (multi-head/FFN)\\
  $B$ & number of FCN/CNN layer\\
  $L$ & number of Enc/Dec layer\\
  $n_h$ & number of attention head\\
  $L_d = 2 l_d + 1$ & length of Enc-D local block\\
  $r_d$ & dropout rate\\
  $Q, K, V$ & query, key, and value in MSA\\
  $L_q, L_k$ ($L^\star$) & subspace lengths of $Q$ and $K, V$\\
  $M, M_G, M_D, M_z$ & attention masks\\
  $\bm{H}_G, \bm{H}_D, \bm{H}_S, \bm{H}_T$ & intermediate inputs\\
  $\bm{P}_G, \bm{P}_D, \bm{P}_S, \bm{P}_T$ & intermediate outputs\\
  \bottomrule
  \end{tabular}
  \endgroup}
  \label{tbl_nota}
\end{table}

\section{Details of Data Sets} \label{apx_data}
\begin{itemize}
  \item Crowd flow prediction: two data sets -- Taxi-NYC and Bike-NYC, are used in the crowd flow prediction. Taxi-NYC is obtained from NYC-TLC, and Bike-NYC is obtained from Citi-Bike. They contain 60 days of trip records, in which the locations and times of the start and the end of a trip is included. We use the first 40 days as training data and the rest 20 days as test data. We also prepare the external information, including rainfall, snowfall, maximum, minimum, average temperatures, and holiday, for ST-ResNet, DMVST-Net, and DSAN. The usages of external information are different according to different methods, which can be referred to in the corresponding papers. Since these data sets are frequently adopted in crowd flow prediction researches, we adopt the general settings, including gird size, time interval, and thresholds, as in the previous works \cite{stdn}.
  \item Service utilization prediction: In this task, we use the CTM data provided by CTM Macau. It contains 45-day records of cellular data service, in which the duration and provider (cell station) location of each service are included. We split the 45-day data set into 30 and 15 days for training and testing. Since the corresponding external information is not available, it is omitted by all methods in this task. We set the grid size to $100m \times 100m$, which is the average coverage of a cell station, and the time interval to $15\ mins$, which is the maximum duration of one record. The thresholds for the two features are set to $60/10$ to filter out idle grids.
\end{itemize}

During the training, we randomly select 20\% of the training data as the validation set. The Numpy files after preprocessing are provided together with our source code.

\begin{table}[ht]
  \centering
  \caption{Details of data. Futures: number of prediction time steps. Thresholds: evaluation thresholds; if the ground truths are less than the thresholds, that instance will be ignored by the RMSE/MAPE. Notice that the max, mean, and standard deviation (Std) are acquired based on the thresholds as well.}
  \resizebox{\linewidth}{!}{
  \begingroup
  \renewcommand{\arraystretch}{1.1}
  \begin{tabular}{ c|c|c|c }
  \hline \hline
  Data sets & Taxi-NYC & Bike-NYC & CTM \\
  \hline
  Map size & $16 \times 12$ & $14 \times 8$ & $20 \times 21$ \\
  \hline
  Grid size & $1km \times 1km$ & $1km \times 1km$ & $0.1km \times 0.1km$ \\
  \hline
  Time interval & 30 mins & 30 mins & 15 mins \\
  \hline
  Time step/day & 48 & 48 & 96 \\
  \hline
  \multirow{2}{*}{Features} & inflow/ & inflow/ & duration (mins)/ \\
  & outflow & outflow & request number\\
  \hline
  Futures & 12 (6 hrs) & 12 (6 hrs) & 12 (3 hrs) \\
  \hline
  Max & 1409/1518 & 262/274 & 17k/3450 \\
  \hline
  Mean & 114.0/146.1 & 33.1/32.6 & 772.1/129.2 \\
  \hline
  Std & 141.6/167.2 & 26.7/26.9 & 1040.7/152.2 \\
  \hline
  Thresholds & 10/10 & 10/10 & 60/10 \\
  \hline
  \multirow{2}{*}{Time Span} & 1/1/2016 - & 8/1/2016 - & 10/11/2018 - \\
  & 2/29/2016 & 9/29/2016 & 11/14/2018\\
  \hline
  Total records & 28.1 million & 3.8 million & 63.6 million \\
  \hline \hline
  \end{tabular}
  \endgroup}
  \label{tbl_datasets}
\end{table}

\subsection{Data Preprocessing}
Min-Max normalization is utilized to convert the data to a scale of [0, 1] during the training and testing. When predicting the next $t_1\ to\ t_{12}$ times steps with DSAN, we follow the rule introduced in \cite{stresnet} and select the data from the same time steps of $t_1$ in previous $N_w$ weeks, which are $t_{1 - (N_w \times a \times 7)}$ to $t_{1 - (1 \times a \times 7)}$, previous $N_d$ days, which are $t_{1 - (N_d \times a)}$ to $ t_{1 - (1 \times a)}$, and previous $N_p$ intervals, which are $t_{1 - N_p}$ to $t_0$, to construct the spatial-temporal input $\bm{X}$, where $a$ is the number of time step per day. In our work, we use $N_w = 1$, $N_d = 3$, and $N_p = 1$. Similar strategies are also employed in the deep learning baselines with settings optimized by their authors. During the calculations of RMSE and MAPE, we exclude all instances whose ground truths are less than the thresholds introduced in Table~\ref{tbl_datasets} for all methods.

\begin{table}[ht]
  \centering
  \caption{The training times. The reported numbers are the approximated averages after running each method for 10 times. The evaluations are performed on the same machine indicated in Appendix~\ref{apx_hyper}.}
  % \resizebox{\linewidth}{!}{
  \begingroup
  \renewcommand{\arraystretch}{1.2}
  \begin{tabular}{ c | c c c }
  \toprule
  & Taxi-NYC & Bike-NYC & CTM \\
  \midrule
  DSAN & 4.4 hrs & 2.6 hrs & 8.3 hrs \\
  \midrule
  MLP & 18 mins & 12 mins & 36 mins \\
  LSTM & 1.5 hrs & 39 mins & 2.1 hrs \\
  GRU & 1.5 hrs & 41 mins & 2 hrs \\
  ConvLSTM & 18 hrs & 15 hrs & 31 hrs \\
  ST-ResNet & 2.8 hrs & 1.7 hrs & 5.3 hrs \\
  DMVST-Net & 14 hrs & 8 hrs & 23 hrs \\
  STDN & 16 hrs & 9 hrs & 28 hrs \\
  \bottomrule
  \end{tabular}
  \endgroup
  \label{tbl_traintime}
\end{table}

\section{Implementations of Baselines} \label{apx_baselines}
\begin{itemize}
  \item \textbf{(1) MLP}: we evaluated the performance of MLP over different hidden unit number -- \{32, 64, 128\}, and empirically observed that 64 achieves the best performance.
  \item \textbf{(2) LSTM}: for LSTM, we select the number of hidden unit from \{32, 64, 128\} to evaluate the performance of LSTM. Empirically, 64 hidden units achieve the best performance.
  \item \textbf{(3) GRU}: hyperparameter setting is identical to LSTM.
  \item \textbf{(4$\sim$7) Deep learning methods}: source codes are obtained from authors' GitHubs.
\end{itemize}
  
We implement the basic neural network methods (\textbf{1}$\sim$\textbf{3}) using Tensorflow and obtain the deep learning baselines (\textbf{4}$\sim$\textbf{7}) from the authors' GitHubs. We tune these models on the validation set using early stopping with 5-epoch window size. Since most of the deep learning methods (e.g., ST-ResNet, DMVST-Net, and STDN) are originally studied for crowd/traffic flow prediction using similar data sets, their original hyperparameters achieve the best performances. As they also consider external information in the predictions, we provide them with the required data, such as weather, holiday, and crowd transition volume. The hyperparameters of these deep learning methods are tuned to acquire the best performances in service utilization prediction. Notice that, since they focus on crowd/traffic flow prediction or precipitation prediction, we modify the input shape of the CTM data according to their frameworks. All baselines use Adam as the optimizers with the default parameters or the ones suggested by their authors. The training times of DSAN and other baselines are shown in Table~\ref{tbl_traintime}. Notice that DSAN is implemented to work with multiple GPUs, while the other deep learning methods are constructed to work on a single GPU. Since their source codes are unformatted and difficult to reconstruct, we have not evaluated the multi-GPU training times of the deep learning baselines.

\section{Hyperparameters} \label{apx_hyper}
We tune DSAN on the validation sets, and observe that $L = 3$, $d = 64$, $n_h = 8$, $B = 3$, $d_f = 256$, $l_d = 3$, and $r_d = 0.1$ achieve the best performances in both tasks. Adam optimizer is used with warm-up learning rate as introduced in \cite{transformer}. Being trained on a machine with 4 NVIDIA RTX2080Ti GPUs, it takes around 4.4 hours for Taxi-NYC, 2.6 hours for Bike-NYC, and 8.3 hours for CTM with 512 as the batch size.

We also evaluate various hyperparameter settings to measure the corresponding performance on Taxi-NYC data to demonstrate the effects of different hyperparameters. Each result in Table~\ref{tbl_hyper} is acquired by averaging the RMSE outputs after running ten times on each setting.
 
As shown in the table, a bigger model is not always better. When a model contains more parameters, it is more vulnerable to over-fitting. The number of attention head $n_h$ also has its appropriate range, which can neither be too low nor too high. Furthermore, we observe that dropout is very helpful in alleviating the over-fitting.
   
\begin{table}[ht]
  \centering
  \caption{Variations on the DSAN architecture. Unlisted values are identical to those of the base set. Taxi-NYC data is used for this evaluation, and all variations perform short-term prediction only.}
  \resizebox{\linewidth}{!}{
  \begin{tabular}{ c | c c c c c c c | c}
  \hline
  & $L$ & $d$ & $d_f$ & $n_h$ & $B$ & $l_d$ & $r_d$ & Inflow/Outflow \\
  \hline
  base & 3 & 64 & 256 & 8 & 3 & 3 & 0.1 & \textbf{17.65}/\textbf{22.86}\\
  \hline
  \multirow{3}{*}{$L$} & 2 & & & & & & & 18.22/23.46 \\
  & 4 & & & & & & & 17.62/22.89\\
  & 6 & & & & & & & 17.93/23.18\\
  \hline
  \multirow{2}{*}{$d$}& & 32 & & & & & & 17.85/23.02 \\
  & & 128 & & & & & & 17.79/22.95\\
  \hline
  \multirow{2}{*}{$d_f$} & & & 128 & & & & & 17.83/23.09\\
  & & & 512 & & & & & 18.00/23.22\\
  \hline
  \multirow{4}{*}{$n_h$}& & & & 1 & & & & 17.83/23.11\\
  & & & & 4 & & & & 17.87/23.15\\
  & & & & 16 & & & & 17.89/23.21\\
  & & & & 32 & & & & 17.96/23.29 \\
  \hline
  \multirow{2}{*}{$B$}& & & & & 1 & & & 18.42/23.71\\
  & & & & & 5 & & & 17.97/23.27 \\
  \hline
  \multirow{2}{*}{$r_d$} & & & & & & & 0.0 & 19.15/24.42 \\
  & & & & & & & 0.2 & 17.71/22.89 \\
  \hline \hline
  \end{tabular}}
  \label{tbl_hyper}
\end{table}

\end{document}